\documentclass[prd, aps, nofootinbib,preprintnumbers, showpacs,superscriptaddress,twocolumn]{revtex4}

\usepackage{latexsym}
\usepackage{amssymb}
\usepackage{amsfonts}
\usepackage{amsmath}
\usepackage{bm}
\usepackage[dvips]{graphicx}
\usepackage{color}
\usepackage{times}
\usepackage{units}

\newcommand{\cf}{\textit{cf.}~}
\newcommand{\ie}{\textit{i.e.,}~}

\newcommand{\pls}{~~~}

\newcommand{\code}[1]{\texttt{#1}}

\allowdisplaybreaks

\begin{document}

\title{Dynamical damping terms for symmetry-seeking shift conditions}

\date{\today}
\label{firstpage}

\author{Daniela Alic}
\affiliation{
  Max-Planck-Institut f\"ur Gravitationsphysik,
  Albert-Einstein-Institut,
  Potsdam-Golm, Germany
}

\author{Luciano Rezzolla}
\affiliation{
  Max-Planck-Institut f\"ur Gravitationsphysik,
  Albert-Einstein-Institut,
  Potsdam-Golm, Germany
}
\affiliation{
  Department of Physics and Astronomy,
  Louisiana State University,
  Baton Rouge, LA, USA
}

\author{Ian Hinder}
\affiliation{
  Max-Planck-Institut f\"ur Gravitationsphysik,
  Albert-Einstein-Institut,
  Potsdam-Golm, Germany
}

\author{Philipp M\"osta}
\affiliation{
  Max-Planck-Institut f\"ur Gravitationsphysik,
  Albert-Einstein-Institut,
  Potsdam-Golm, Germany
}

\begin{abstract}
Suitable gauge conditions are fundamental for stable and accurate
numerical-relativity simulations of inspiralling compact binaries. A
number of well-studied conditions have been developed over the last
decade for both the lapse and the shift and these have been
successfully used both in vacuum and non-vacuum spacetimes when
simulating binaries with comparable masses. At the same time, recent
evidence has emerged that the standard ``Gamma-driver'' shift
condition requires a careful and non-trivial tuning of its parameters
to ensure long-term stable evolutions of unequal-mass binaries. We
present a novel gauge condition in which the damping constant is
promoted to be a dynamical variable and the solution of an evolution
equation. We show that this choice removes the need for special tuning
and provides a shift damping term which is free of instabilities in
our simulations and dynamically adapts to the individual positions and
masses of the binary black-hole system. Our gauge condition also
reduces the variations in the coordinate size of the apparent horizon
of the larger black hole and could therefore be useful when simulating
binaries with very small mass ratios.
\end{abstract}

\maketitle

\section{Introduction}
\label{introduction}

Five years after the first demonstrations~\cite{Pretorius:2005gq,
  Campanelli:2005dd, Baker:2005vv} that the numerical solution of the
inspiral and merger of binary black holes (BBHs) was within the
technical and computational capabilities of many numerical-relativity
groups, our understanding of this process has expanded beyond the most
optimistic predictions. Numerical-relativity simulations of black-hole
binaries have been performed in a large region of the possible
space of parameters (see~\cite{Hinder:2010vn,Campanelli:2010ac} for
two recent reviews). In addition, the results of these simulations
have been exploited on several fronts. In gravitational wave data analysis, they have been used to produce
template banks that increase the distance reach of
detectors~\cite{Ajith:2007qp, Ajith:2007kx, Ajith:2007xh} and to aid
the calibration of search pipelines \cite{Aylott:2009ya, Farr:2009pg,
  Santamaria:2009tm}. In astrophysics they have been used to determine the properties of
the final black hole (BH) of a BBH inspiral and merger (see~\cite{Rezzolla:2008sd, Barausse:2009uz,
  vanMeter:2010md} and references therein for some recent work) and
hence assess the role that the merger of supermassive BHs plays in the
formation of galactic structures~\cite{Berti2008,Fanidakis:2009}. In
cosmology they have been used to study the electromagnetic counterparts to the merger of
supermassive BH binaries and hence deduce their
redshift~\cite{Palenzuela:2009hx, Moesta:2009, Bode:2009mt,
  Farris:2009mt}.

Despite this extensive and rapid progress, there are portions of the
space of the parameters that still pose challenges for numerical
simulations, in particular those involving binaries with BHs that are
maximally spinning (but see~\cite{Dain:2008} for some progress in this
direction) or with small mass ratios $q\equiv M_2/M_1 \leq 1$. This
latter problem is potentially a rather serious one since the
computational costs scale in general quadratically with the inverse of
the mass ratio of the system. The inspiral timescale is inversely
proportional to $q$ (the smaller BH spends more orbits per frequency
interval during its inspiral onto the larger BH) and the timestep
limit in explicit numerical schemes also decreases linearly with the
BH size and hence inversely proportionally with $q$.

While some progress has been made recently when simulating binaries
with mass ratios as small as $q=1/10$~\cite{Gonzalez:2008,
  Lousto:2010tb}, it is clear that some significant technical changes
are needed in order to tackle mass-ratios which are much smaller. One
of these changes may consist in adopting implicit numerical schemes in
which the timestep limitation is set uniquely by the truncation error
and not by the smallest spacing of the spatial numerical grid. Another improvement
could come from the use of better spatial gauge conditions that, by
better adapting the coordinates to the different curvatures of the
spatial slice, may reduce the numerical error and hence the
computational cost. Recent numerical simulations have revealed that the
standard ``Gamma-driver'' shift condition~\cite{Alcubierre02a}
requires a careful tuning of its parameters to ensure
long-term stable BBH evolutions when considering unequal-mass
binaries. Work to alleviate some of these problems has been recently
started~\cite{Mueller:2009jx, Mueller:2010bu, Schnetter:2010cz} and
has so far concentrated on adapting the damping term $\eta$ in the
shift condition (see Sect.~\ref{setup} for a detailed discussion of
the gauge and of the damping term) to better suit the uneven
distribution of local curvature on the spatial hypersurface as the
binary evolves. In practice, while a constant value of $\eta$ has
worked well for comparable-mass binaries, the investigations reported
in~\cite{Mueller:2009jx, Mueller:2010bu, Schnetter:2010cz} have
suggested the use of damping factors that have a spatial dependence
adapted to the location of the BHs.
The first non-constant prescription for $\eta$ \cite{Mueller:2009jx}
used an expression which adapts to the mass of each BH via the
conformal factor, but it was found to lead to large errors at mesh
refinement boundaries \cite{Mueller:2010bu} and so was replaced by a
simpler analytical form containing constant parameters which need to
be tuned to the masses of the BHs.
At present it is not clear whether the tuning made with mass
ratios $q\gtrsim 0.25$ will be effective also for much smaller mass
ratios.

In this paper we propose a different approach to the problem of a
dynamical damping term for symmetry seeking shift conditions and
promote $\eta$ to be a fully dynamical variable, as has proven very effective
for the shift vector and for the
$\tilde \Gamma$ variables.
The resulting gauge condition dynamically adapts to the individual
positions and masses of the BBH system without the need for special
tuning and remains well-behaved (\ie smooth and bounded) at all
times.
Considering three different options for the source term in the
evolution equation for $\eta$, we show that an expression which is
very simple to implement leads to numerical errors which are
comparable to or smaller than the constant $\eta$ case.  Furthermore,
this choice reduces the dynamics in the coordinate size of the
apparent horizon of the smaller BH and could therefore be useful when
simulating binaries with very small mass ratios.

The structure of the paper is as follows. In Sect.~\ref{setup} we
summarise the numerical infrastructure and the mathematical setup
used in our simulations, while Sect.~\ref{equations} is dedicated to a
review of the slicing and spatial gauge conditions and to the
discussion of our novel approach. Sections~\ref{asbh}
and~\ref{abbh} are dedicated to the discussion of the results
of applying the new gauge to simulations of single nonspinning
BHs (Sect.~\ref{asbh}) and to systems with BHs having
either equal or unequal masses (Sect.~\ref{abbh}). Finally, the
conclusions and the prospects for future work are detailed in
Sect.~\ref{conclusions}.  We use a spacelike signature $(-,+,+,+)$ and
a system of units in which $c=G=M_\odot=1$.

\section{Numerical Setup}
\label{setup}

The numerical setup used in the simulations presented here is the same
one discussed in~\cite{Pollney:2007ss} and more recently applied to
the \code{Llama} code described in~\cite{Pollney:2009yz}. The latter
makes use of higher-order finite-difference algorithms (up to $8$th
order in space) and a multi-block structure for the outer
computational domain, which allows one to move the outer boundary to a
radius where it is causally disconnected from the binary. We refer the
reader to the papers above for details and here simply note that we
solve the Einstein equations in vacuum with a conformal and traceless
formulation of the equations, in which the conformal factor
has been redefined as $W \equiv [({\rm det} (\gamma_{ab})]^{-1/6}$, or
in terms of the metric $\tilde{\gamma}_{ab} = W^{2} \gamma_{ab}$. The
corresponding evolution equation is therefore
\begin{eqnarray}
\partial_{t} W - \beta^{i} \partial_{i} W = \frac{1}{3} W \alpha K 
- \frac{1}{3} W \partial_{i} \beta^{i}\,,
\end{eqnarray}
where $K$ is the trace of the extrinsic curvature
(see~\cite{Pollney:2007ss} and~\cite{Pollney:2009yz} for details on
our specific implementation).

\begin{figure}
\includegraphics{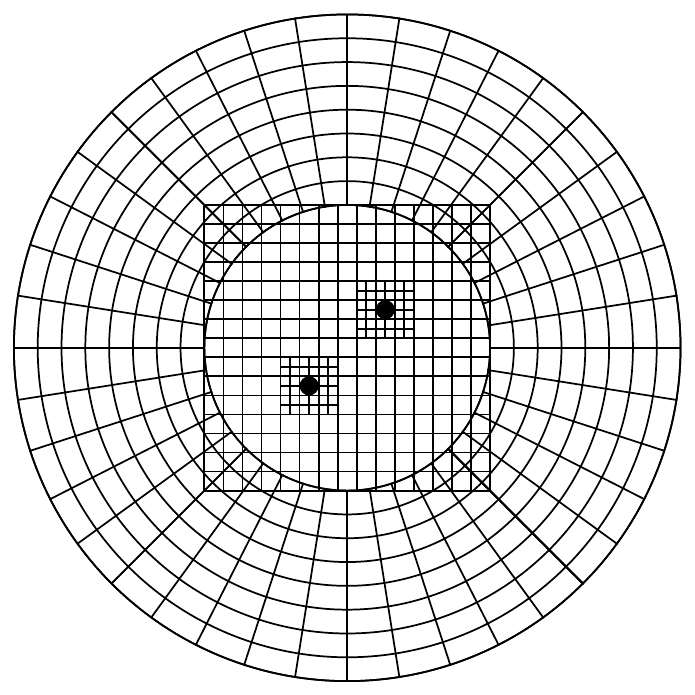}
\caption{Schematic diagram of a typical \texttt{Llama} grid setup in
  the $(x,y)$ plane. Note the inner Cartesian grid with box-in-box AMR
  which is joined to a $6$-patch multiblock structure (only $4$ of
  these patches are shown in the $(x,y)$ plane).}
\label{llama_scheme}
\end{figure}

The computational infrastructure of the \code{Llama} code is based on
the \texttt{Cactus} framework \cite{Goodale02a,Cactusweb} and the
\texttt{Carpet} \cite{Schnetter-etal-03b,carpetweb} mesh-refinement
driver, and implements a system of multiple grid patches with data
exchanged via interpolation~\cite{Pollney:2009yz}. We use a central
cubical Cartesian patch containing multiple levels of adaptive mesh
refinement, with higher resolution boxes tracking the location of each
BH, \ie ``moving boxes''. This is surrounded by $6$ additional patches
with the grid points arranged in a spherical-type geometry, with
constant angular resolution to best match the resolution requirements
of radially outgoing waves.  This allows us to evolve out to very
large radii at a tiny fraction of the computational cost which would
be necessary to achieve the same resolution with a purely Cartesian
code.  We use one patch for each of the $\pm x$, $\pm y$ and $\pm z$
axes, which leads to an inner spherical interpatch boundary and to a
spherical outer boundary. The latter can be placed at very large
distances and we choose it to be it causally disconnected from the
surfaces on which we compute the waveforms for the duration of the
simulation.  Figure~\ref{llama_scheme} shows a schematic diagram of a
typical \texttt{Llama} grid setup in the $(x,y)$ plane. Note the inner
Cartesian grid with moving-boxes which is
joined to a $6$-patch multiblock structure (only $4$ of these patches
are shown in the $(x,y)$ plane). The detailed grid structure used in
each run is listed in Table \ref{tab:grid}, and for the purpose of
comparison, the resolution of each simulation is indicated by the grid
spacing $h_0/M$ of the coarsest Cartesian grid.  The unit $M$ is
chosen such that each BH has mass $0.5 M$ in both the single and
binary BH cases.  In all cases the coarsest resolution is also equal
to the radial spacing in the angular patches.

\begin{table*}
 \begin{tabular}{l|ccccc|cccccc}
\hline
\hline
~Configuration & $h_0/M$          & $N_\mathrm{ang.}$ & $R_\mathrm{in}/M$ & $R_\mathrm{out}/M$ & $N_\mathrm{lev.}$ & $r_{\text{l}}/M$ \\
\hline
~single BH           & $0.96$             & $21$          & $39.36$   & $400.00$  & $6$    & $(12, 6, 3, 1.5, 0.6)$ \\
~BBH, $q = 1$   & $0.96$            & $21$          & $39.36$ &
$1980.48$ & $6$    & $(12, 6, 3, 1.5, 0.6)$ \\
~BBH, $q = 1/4$~   &~~$(0.8, 0.96, 1.12)$~~  & ~~$(23, 27, 33)$~~  & $49.92$   & $2545.92$ & $(6, 8)$ & ~~$(12, 6, 3, 1.5, 0.8, 0.4, 0.2)$~~ \\
\hline
\hline
\end{tabular}
\caption{Numerical grid parameters of each BH configuration studied.
  $h_0$ is the grid spacing on the coarsest Cartesian grid, which is
  equal in all cases to the radial grid spacing in the angular
  patches. $N_\mathrm{ang.}$ is the number of cells in the angular
  directions in the angular patches. $R_\mathrm{in}$ and
  $R_\mathrm{out}$ are the inner and outer radii of the angular
  patches. $N_\mathrm{lev.}$ is the number of refinement levels
  (including the coarsest) on the Cartesian grid, and $r_{\text{l}}$
  indicates that a cubical refinement box of side $2 r_{\text{l}}$ is
  centred on the BH on level ``l'', with level 0 being the coarsest
  (in the case that the boxes overlap, they are replaced with a single
  box enclosing the two).  The unit $M$ is chosen such that each BH has
  mass $0.5 M$ in both the single and binary BH cases.}
\label{tab:grid}
\end{table*}

\section{Gauge conditions and Dynamical Damping Term}
\label{equations}

As mentioned in the introduction, for the formulation of the Einstein
equations we adopt, the use of suitable gauge conditions was the last
obstacle to overcome in order to obtain long-term stable simulations
of BBHs~\cite{Campanelli:2005dd, Baker:2005vv}. In what is now the
standard {\em moving-puncture} recipe, the lapse $\alpha$ is evolved
using a singularity-avoiding slicing condition from the $1+\log$
family~\cite{Bona94b}
\begin{equation}
\partial_{t} \alpha - \beta^{i} \partial_{i} \alpha = -2 \alpha K\,,
\end{equation}
while the shift vector $\beta^{i}$ is evolved using the hyperbolic
Gamma-driver condition~\cite{Alcubierre02a}
\begin{eqnarray}
\label{dtbeta}
\partial_{t} \beta^{a} - \beta^{i} \partial_{i} \beta^{a} &=& 
\frac{3}{4} B^{a}, \\
\label{dtB}
\partial_{t} B^{a} - \beta^{i} \partial_{i} B^{a} 
&=& \partial_{t} \tilde{\Gamma}^{a} - 
\beta^{i} \partial_{i} \tilde{\Gamma}^{a} - \eta B^{a}\,.
\end{eqnarray}
We recall that the Gamma-driver shift condition is similar to the
Gamma-freezing condition $\partial_t \tilde\Gamma^k=0$ which, in
turn, is closely related to the minimal distortion shift
condition~\cite{Smarr78b}. The differences between these two
conditions involve the Christoffel symbols and are basically due to
the fact that the minimal distortion condition is covariant, while the
Gamma-freezing condition is not (see the discussion
in~\cite{Baiotti04}).

The coefficient $\eta$ of the last term in~\eqref{dtB} is usually
referred to as the damping term and plays a fundamental role in our
investigation. It was originally introduced to avoid strong
oscillations in the shift and experience has shown that by tuning its
value, it is possible to essentially ``freeze'' the evolution of the
system at late times~\cite{Alcubierre2003:hyperbolic-slicing}. 
In simulations of inspiralling compact binaries this
damping term is typically set to be constant in space and time and
equal to $2/M$ for BBHs and equal to $1/M$ for binaries of neutron
stars, where $M$ is the sum of the masses of the BHs or neutron stars~\cite{Baiotti08, Rezzolla:2010}.
Similar values have also been
shown to yield stable evolutions in the case of mixed binaries with
mass ratios $q\simeq 1/6$~\cite{Loeffler06a}.
While this choice works well for binaries with comparable masses, a
simple dimensional argument shows that it will cease to be a good one
for binaries with unequal masses.  Since the dimension of $\eta$ is
inverse mass, and the relevant mass is the mass of each individual BH,
as the BH mass decreases a larger value of $\eta$ will be needed to
maintain a similar damping effect.  For an unequal mass system, this
is impossible with a constant $\eta$.

\subsubsection{Position-Dependent Damping Term}

To overcome the limitations imposed by a constant-in-space damping
term, various recipes have been proposed recently in the literature. A
first suggestion was presented in Ref.~\cite{Mueller:2009jx}, where
the damping term was specified by the function
\begin{eqnarray}\label{DampD1}
\eta_{\text{MB}}(r) = R_{0} \frac{\sqrt{\tilde{\gamma}^{ij} 
\partial_{i}W \partial_{j}W}}{ (1 - W)^{2} }\,,
\end{eqnarray}
where $\tilde{\gamma}^{ij}$ is the inverse of the conformal 3-metric
and $R_{0}$ is a dimensionless constant chosen in such a way that
$R_{0}M_\mathrm{BH}$ corresponds to the Schwarzschild radial coordinate for the
stationary state, \ie $R_{0} \simeq
1.31241$~\cite{Hannam:2008sg}. However, as discussed subsequently by
the same authors \cite{Mueller:2010bu}, expression~\eqref{DampD1}
leads to sharp features (spikes) which can produce coordinate drifts
and affect the stability of the simulations. To remove these
drawbacks, alternative forms were suggested that read
respectively~\cite{Mueller:2010bu}
\begin{equation}
\eta_{\text{MGB}} = A + \frac{C_1}{1 + w_1\,(\hat{r}_1^2)^n} + \frac{C_2}{1 +
    w_2\,(\hat{r}_2^2)^n}\,,
  \label{eq:etaS6}
\end{equation}
and
\begin{equation}
\eta_{\text{MGB}} = A + C_1e^{-w_1\,(\hat{r}_1^2)^n} + C_2e^{-w_2\,(\hat{r}_2^2)^n}\,,
    \label{eq:etaS5}
\end{equation}
where $w_1$ and $w_2$ are positive parameters chosen to change the
width of the functions based on the masses of the two BHs.  The power
$n$ is a positive integer which determines the fall-off rate, while
the constants $A$, $C_1$, and $C_2$ are chosen to provide the desired
values of $\eta$ at the punctures and at infinity. Finally, the
dimensionless radii $\hat{r}_1$ and $\hat{r}_2$ are defined as
$\hat{r}_i = {|\vec{r}_i-\vec{r}|}/{|\vec{r}_1-\vec{r}_2|}$, where $i$
is either $1$ or $2$, and $\vec{r}_i$ is the position of the $i$-th
BH. In addition to the new
prescriptions~\eqref{eq:etaS6}--\eqref{eq:etaS5}, which provide
appropriate values both near the individual punctures and far away
from them with a smooth transition in between, there is evidence that
they also lead to a smaller truncation error. In particular, when
examined for $w_1=w_2=12$, $n=1$, the waveforms produced when using
Eq.~(\ref{eq:etaS6}) showed less deviation with increasing resolution
than using a constant $\eta$. Similar results were found when using
Eq.~(\ref{eq:etaS5}), leading the authors to the conclusion that the
preferred definition for the damping term is Eq.~(\ref{eq:etaS6})
because it is computationally less expensive.

It was shown recently \cite{Schnetter:2010cz} for the Gamma-driver
shift condition that there is a stability limit on the time step size
which depends on $\eta$.  This limitation comes from the time
integrator only and is not dependent on spatial resolution.  One of
the proposed solutions to this problem is to taper $\eta$ with a
functional dependence of the type
\begin{equation}
\label{DampE1}
\eta_{\text{S}}(r) = \eta_0\frac{R^{2}}{r^{2} + R^{2}}\,, 
\end{equation}
where $r$ is the coordinate distance from the centre of the BH and $R$
is the radius at which one makes the transition between an inner
region where $\eta_{\text{S}}$ is approximately equal to $\eta_0$ and
an outer region where $\eta_{\text{S}}$ gradually decreases to zero.
We find that this form does indeed help in removing potential
instabilities and, as we will discuss in the following section, can be
used also for a fully dynamical definition for the damping term.

\subsubsection{Evolved Damping Term}

While the analytic prescriptions discussed in the previous section
have been shown to be effective when suitably tuned, it is not clear
whether they will be equally effective for different mass ratios, nor
how to choose the parameters without case-by-case tuning.  In view of
this and in order to derive an expression which adapts dynamically in
space and time to the local variations of the shift vector, we have
promoted the damping term to be an evolved variable with the simple
equation
\begin{eqnarray}
\label{DampEvol}
\partial_{t}\eta - \beta^{i} \partial_{i} \eta =
\frac{1}{M}\left(- \eta + S(r)\right) 
\,.
\end{eqnarray}
The function $S(r)$ is a position-dependent source term which can be
chosen freely.  The first term on the right-hand-side is introduced to
induce an exponential decay of the damping term towards $S(r)$ such
that in the steady state, $\eta \to S(r)$. The advective derivative
term $\beta^{i} \partial_{i} \eta$ ensures that the motion of the
punctures, which are locally advected by $\beta^i$, is taken into
account in the driving of $\eta \to S(r)$ (we want to drive $\eta$ to
a specific behaviour in the neighbourhood of the punctures). To better
interpret our suggestion for the dynamical gauge~\eqref{DampEvol}, it
is useful to compare it with a simpler ordinary differential equation
\begin{equation}
\label{toy_ode}
\tau \frac{d \eta}{dt} = - \eta + S(t)\,.
\end{equation}
Setting now $\eta_{_{0}}\equiv \eta(t=0)$, $S_{_{0}}\equiv S(t=0)$,
and assuming that $\tau (d S/dt) \ll 1$ so that it can be neglected at
first order, Eq.~\eqref{toy_ode} would have solution
\begin{equation}
\eta \simeq S + (\eta_{{_0}} - S_{_{0}}) e^{-t/\tau}\,,
\end{equation}
and thus $\eta \to S$ as $t \to \infty$. Although in the case of our
gauge-evolution equation~\eqref{DampEvol}, the source term is
time-dependent and sometimes the time derivative can be rather large,
especially near the punctures, equation~\eqref{toy_ode} is useful to
recognise that the damping term is itself damped and driven to the
solution given by the source function $S$.

\begin{figure*}
\includegraphics[width=8.0cm]{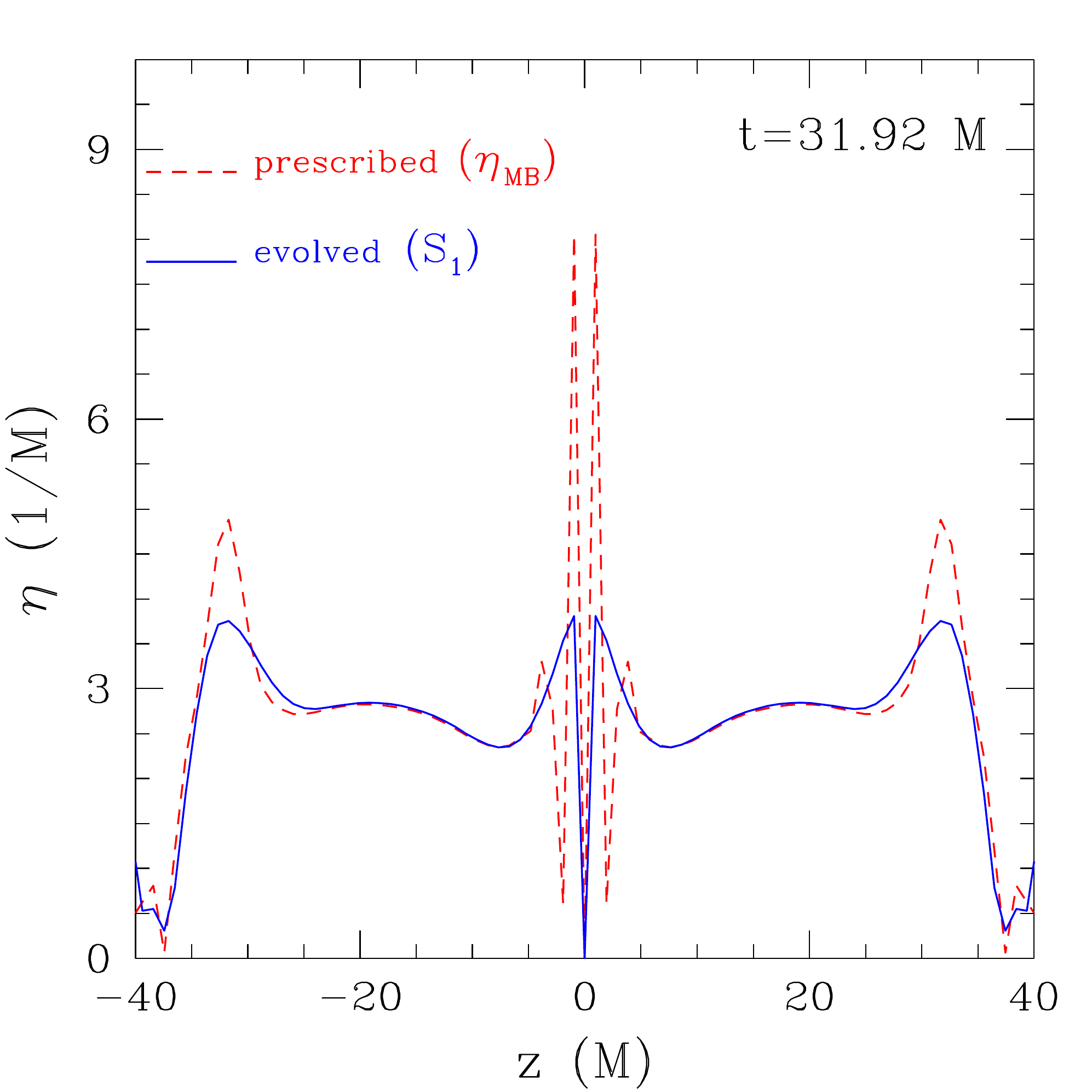}
\hskip 1.0cm
\includegraphics[width=8.0cm]{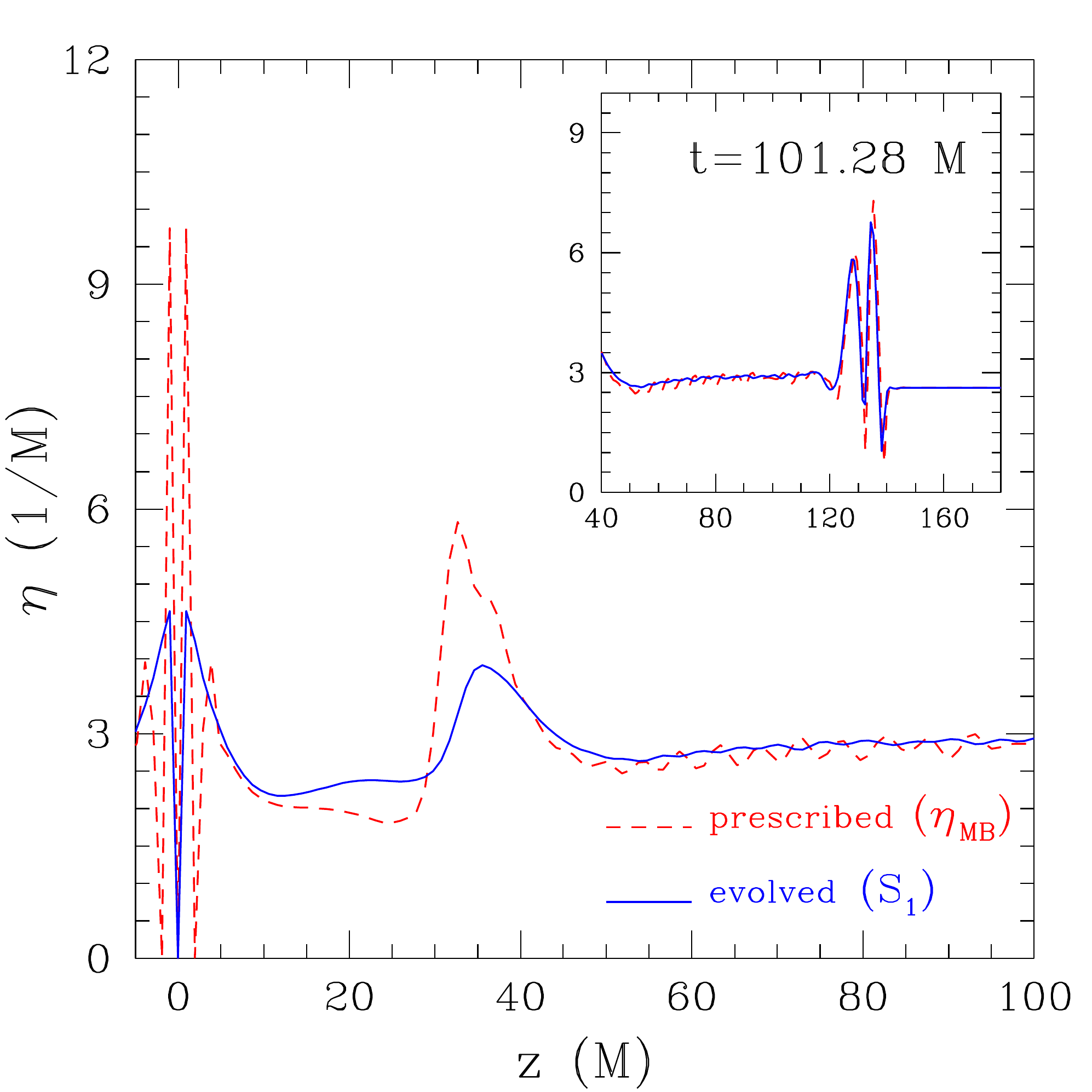}
\caption{Profiles along the $z$-axis of the damping term $\eta$ for an
  isolated Schwarzschild BH. Different lines refer to the case when
  $\eta$ is prescribed using Eq. (\ref{DampD1}) (red dashed line) or
  when evolved in time using Eq. (\ref{DampEvol}) with source given by
  (\ref{SourceD1}) (blue solid line). The left panel refers to an
  earlier time and focuses on the central region of the grid. The
  right panel refers to a later time and shows a larger portion of the
  grid, highlighting that in contrast to the dynamical $\eta$, the
  evolved one leads to smooth profiles at the patch boundary (see
  inset). Note that the outer boundary is at $400\,M$ and cannot be
  responsible for the appearance of the spikes via reflection. }
\label{eDBH}
\end{figure*}

\begin{figure*}
\includegraphics[width=8.0cm]{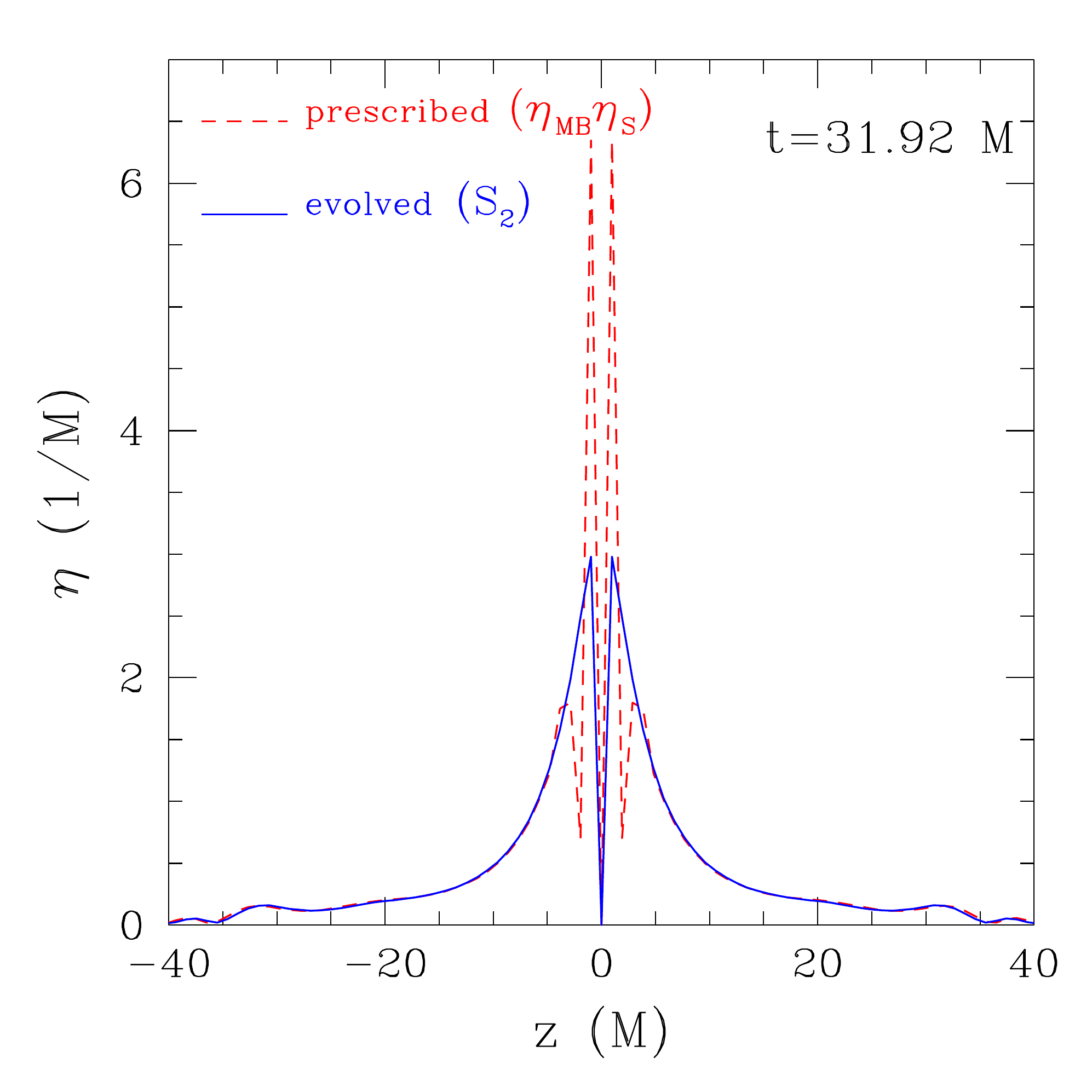}
\hskip 1.0cm
\includegraphics[width=8.0cm]{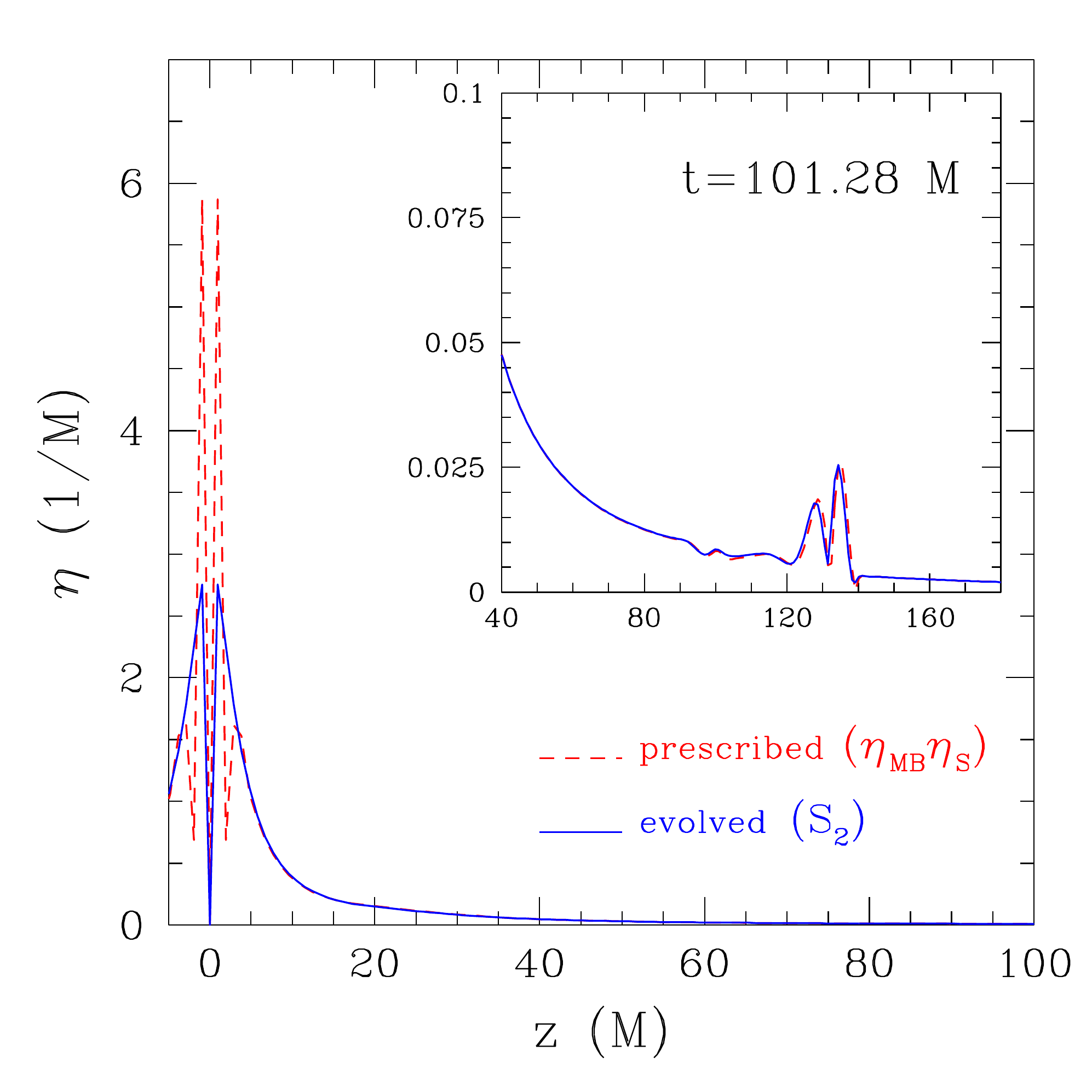}
\caption{The same as in Fig.~\ref{eDBH}, but the prescribed damping
  term Eq. (\ref{DampD1}) is tapered with a function
  Eq. (\ref{DampE1}), while the source term for
  the evolution equation is given by (\ref{SourceED1}). Note that in
  this case the damping term falls off as expected but that in
  the case of the prescribed $\eta_{\text{MB}}(r)\eta_{\text{S}}(r)$, the outgoing gauge
  pulses still produces sharp features near the BH at $z\sim 3\,M$,
  which do not propagate away.  Moreover, sharp features are produced
  as the initial spikes pass through the interpatch boundaries located
  at $z\sim 40\,M$. Smoother profiles can instead be obtained by
  evolving the damping factor.}
\label{eEDBH}
\end{figure*}

The arbitrariness in the form of $S(r)$ is removed in part by the
works discussed in the previous sections and hence a first possible
form is inspired by~(\ref{DampD1}) and thus given by
\begin{equation}
\label{SourceD1}
S_{1}(r) = \eta_{\text{MB}}(r) = R_{0} \frac{\sqrt{\tilde{\gamma}^{ij}
    \partial_{i}W \partial_{j}W}} { (1 - W)^{2} }\,.
\end{equation}
As we will discuss in the next sections, this choice works very well
for single BHs, leading to smoother profiles for $\eta$ and
consequently more stable evolutions. Similarly, another convenient
choice for the source is a combination of expression
(\ref{DampD1}) and (\ref{DampE1}) 
\begin{equation}
\label{SourceED1}
S_{2}(r) = \eta_{\text{MB}}(r) \eta_{\text{S}}(r) 
= \left(R_{0} \frac{\sqrt{\tilde{\gamma}^{ij} 
\partial_{i}W \partial_{j}W}}{ (1 - W)^{2} }\right)
\left(\frac{R^{2}}{r^{2} + R^{2}}\right)\,, 
\end{equation}
This choice ensures that $\eta$ is dynamically adapting in the inner
region due to the $\eta_{\text{MB}}$ factor, while in the outer region
the $\eta_{\text{S}}$ factor ensures minimal dynamics and implements
the suggestion of Ref.~\cite{Schnetter:2010cz} that $\eta \to 0$ at
large radius to avoid the instability there.

\begin{figure*}
\includegraphics[width=8.0cm]{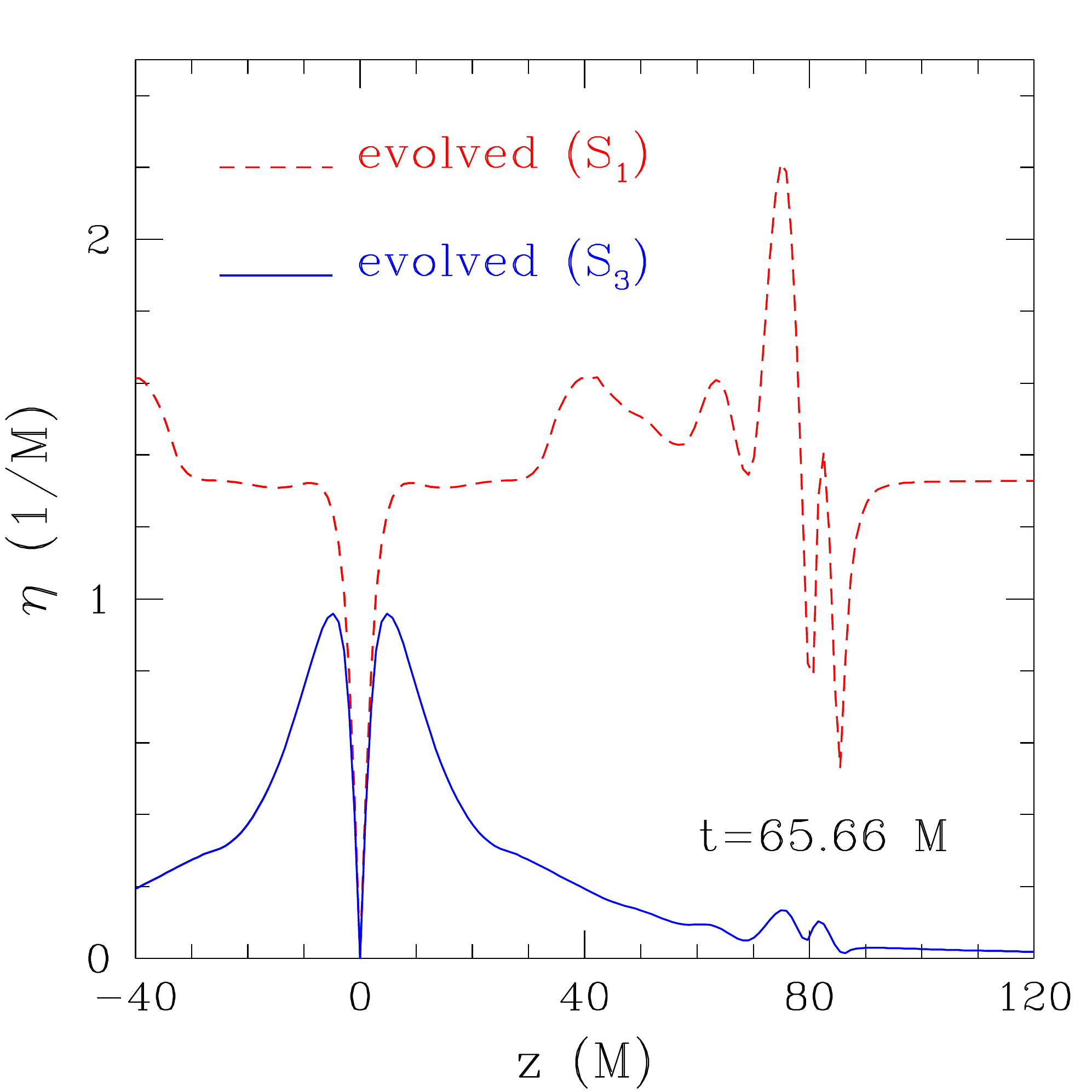}
\hskip 1.0cm
\includegraphics[width=8.0cm]{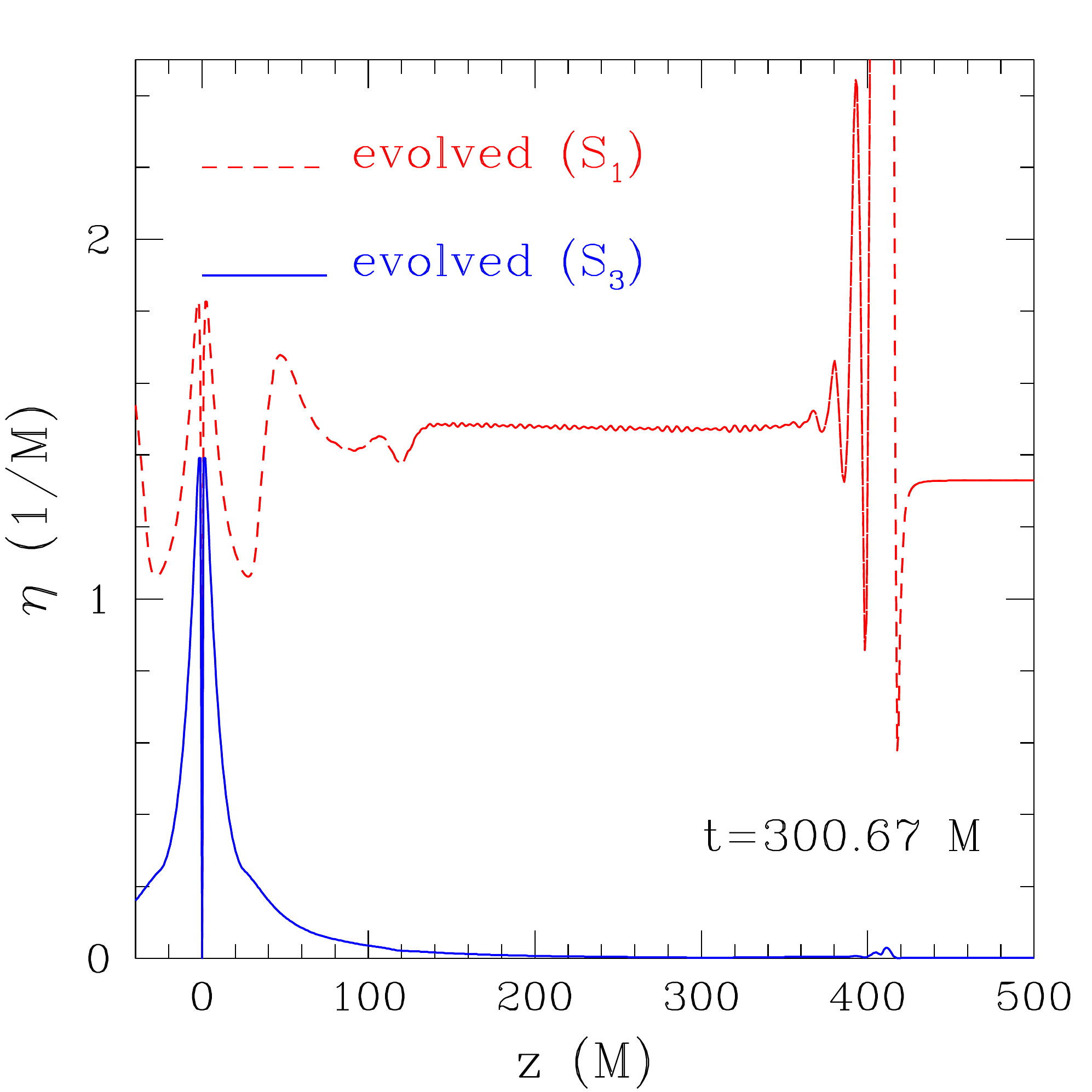}
\caption{Profiles along the $z$-axis of the damping term $\eta$ for an
  \textit{equal-mass} BBH system. Different lines refer to the case
  when $\eta$ evolved with sources given respectively by
  (\ref{SourceD1}) (red dashed line) or by (\ref{SourceED1}) (blue
  solid line). The left and right panel refers to two different times
  and highlight the importance of the fall-off term to avoid
  reflections at mesh boundaries. Note that the outer boundary is at
  $\simeq 2000\,M$ and cannot be
  responsible for the appearance of the spikes via reflection.}
\label{evolBBH}
\end{figure*}

The generalisation of the source term~\eqref{SourceED1} to the case of
a binary system is straightforwardly given by
\begin{equation}
\label{DampE3}
S_{3}(r) = \left(R_{0} \frac{\sqrt{\tilde{\gamma}^{ij} 
\partial_{i}W \partial_{j}W}}{ (1 - W)^{2} }\right)
\left(\frac{R^{2}}{r_{1}^{2} + r_{2}^{2} + R^{2}}\right)\,, 
\end{equation}
where $\vec{r}_1, \vec{r}_2$ are the distances from $\vec{r}$ to the
individual BHs centres.

The following sections will be dedicated to the results of simulations
performed with the evolution equation for the damping
term~\eqref{DampEvol} to study nonspinning single BHs via the
source terms~\eqref{SourceD1}--\eqref{SourceED1}, and to evolve BBHs
via the source term~\eqref{DampE3}. As a final remark we note that
although our dynamical gauge requires the numerical solution of an
additional equation [\ie Eq.~\eqref{DampEvol}], the associated
computational costs are minimal, given that we are evolving a scalar
quantity and that the evolution equation is very similar to those
already implemented within the Gamma-driver condition. More
specifically, the added computational costs range from $1$ to $2\%$
depending on the complexity and length of the simulation (the longer
the simulation, the larger the impact of frequent restarts and thus
the smaller the impact of the additional evolution equation).

\section{Application of the new gauge: single black holes}
\label{asbh}

We next examine the properties of the new dynamical
gauge~\eqref{DampEvol} for the damping factor and present its
advantages when compared with the other position- and mass-dependent
suggestions presented in the previous section.  Specifically, we will
study the $\eta_{\text{MB}}(r)$ and $\eta_{\text{S}}(r)$ prescribed
forms for $\eta$, and compare them with both the $S_{1}(r)$ and
$S_{3}(r)$ variants of the evolved gauge condition~\eqref{DampEvol}.
We will start by considering the simple case of a single nonspinning
puncture.

Fig.~\ref{eDBH} shows a comparison between an evolution using the
prescription $\eta_{\text{MB}}(r)$ provided by Eq.~(\ref{DampD1}) (red
dashed lines) and the new gauge using as source $S_{1}(r)$ from
Eq.~(\ref{SourceD1}) (blue solid lines). In both cases, the value of
the damping factor at the puncture adapts automatically to the mass of
the BH, due to its dependence on the conformal factor.
However, when the damping term is given the form $\eta_{\text{MB}}(r)$,
large spikes develop at the origin at late times and noise
travels outwards as the gauge
settles (\cf left panel of Fig.~\ref{eDBH}). These large noise pulses
leave sharp features in $\eta$ which might lead to coordinate drifts
and eventually affect the stability when simulating BBHs.
On the other
hand, when evolving the damping factor and using $\eta_{\text{MB}}(r)$
as a source, it is possible to avoid any forcing of the
damping term, which is instead always the solution of the dynamical
driver. This, in turn, reduces the noise and ensures long-term
stability of the simulation as shown by the right panel of
Fig.~\ref{eDBH}.

We can further improve the stability properties of the damping factor
in the outer wave region by matching it with a function which drives
it smoothly to zero.  Fig.~\ref{eEDBH} shows a comparison between a BH
evolution using the prescription $\eta_{\text{MB}}(r)\eta_{\text{S}}(r)$ given by
Eqs.~(\ref{DampD1}) and (\ref{DampE1}) (red dashed lines), and the evolved $\eta$ using as
source Eq.~(\ref{SourceED1}) (blue solid lines). Note that in the case
of the prescribed $\eta_{\text{MB}}(r)\eta_{\text{S}}(r)$, the outgoing gauge pulses
still produces sharp features near the BH at $z\sim 3\,M$, which are neither
propagated away nor damped in-place. 
Moreover, sharp features far from the BH continue to be produced as
the initial spikes pass through the interpatch boundaries.

Smoother profiles can instead be obtained by evolving the damping
factor, using $\eta_{\text{MB}}(r)\eta_{\text{S}}(r)$ as a source. This is especially
true near the BH, while large variations but of small amplitude are
still produced as the gauge pulses pass through the interpatch
boundaries. Finally we note that in contrast to all the other
evolution variables in our code, no artificial dissipation is imposed
on the damping term so that the features of its evolution equation can
be better appreciated.

\begin{table*}[ht]
\begin{tabular}{l|cccccc}
\hline
\hline
~Configuration & $m_1/M$ & $m_2/M$ & $x_1/M$ & $x_2/M$ & $P^x_1/M$ & $P^y_1/M$ \\
\hline
~single BH      & $0.5$  & $--$   & $0.0$  & $--$   & $\pls 0.00000000$ & $0.00000000$ \\
~BBH, $q = 1$   & $0.5$  & $0.5$  & $3.5$  & $-3.5$ &     $-0.00335831$ & $0.12369380$   \\
~BBH, $q = 1/4$ & $0.8$  & $0.2$  & $1.6$  & $-6.4$ &     $-0.00104474$ & $0.07293950$  \\
\hline
\hline
\end{tabular}
\caption{Initial data parameters for the configurations studied. Expressed in units of the
  total mass $M$ are: the initial irreducible masses of the BHs
  (obtained by iterating over bare mass parameters) $m_i$, the
  coordinates of the BHs on the $x$-axis $x_i$, the momentum of BH 1
  $P^i_1$ (the momentum of BH 2 is equal and opposite,
  $P^i_2=-P^i_1$). The unit $M$ is chosen such that each BH has
  mass $0.5 M$ in both the single and binary BH cases.}
\label{tab:configs}
\end{table*}

\section{Application of the new gauge: black-hole binaries}
\label{abbh}

As shown in the previous section, our evolved damping factor leads to
smoother profiles and consequently to stable BH evolutions, free of
coordinate drifts.  In this subsection, we study the effect of using
different functional forms for the sources in the evolution
equation~\eqref{DampEvol}.

\subsection{Equal-mass binaries}

We first consider the evolution of an equal-mass nonspinning BH
binary, whose properties can be found in Table~\ref{tab:configs}. For
simplicity we have considered a system with small separation
$D=7\,M$, so that overall the binary performs only about $3$ orbits
before merging and settles to an isolated spinning BH after about
$200\,M$.

Fig.~\ref{evolBBH} shows a comparison of the profile of the damping
term on the $z$-axis when using the evolution
equation~\eqref{DampEvol} and the source term given either by $S_1$
[\cf Eq.~(\ref{SourceD1}); red dashed line], or by $S_3$ [\cf
  Eq.~(\ref{DampE3}); blue solid line]; in both
cases we have set $R=20M$.

\begin{figure*}
\includegraphics[width=8.0cm]{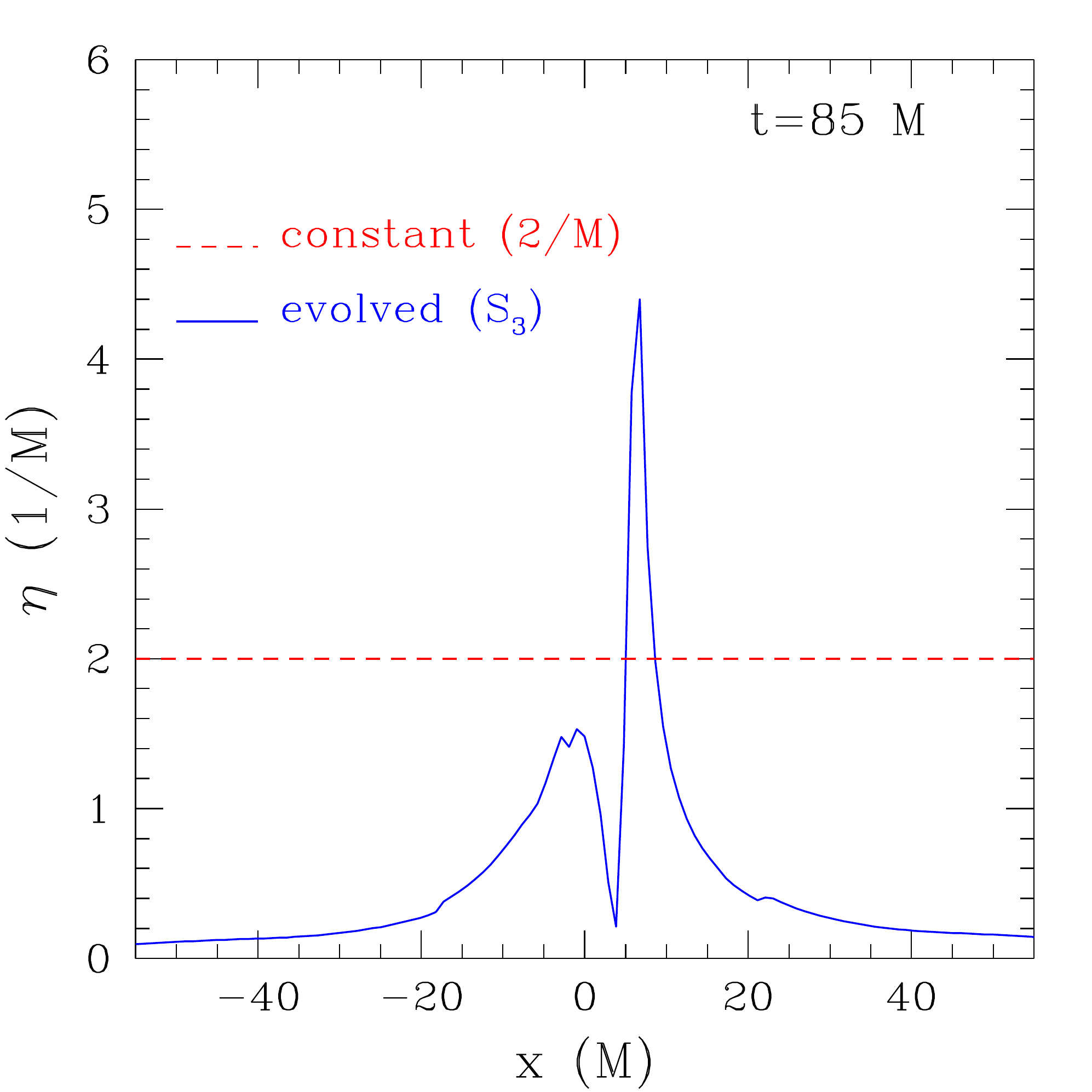}
\hskip 1.0cm
\includegraphics[width=8.0cm]{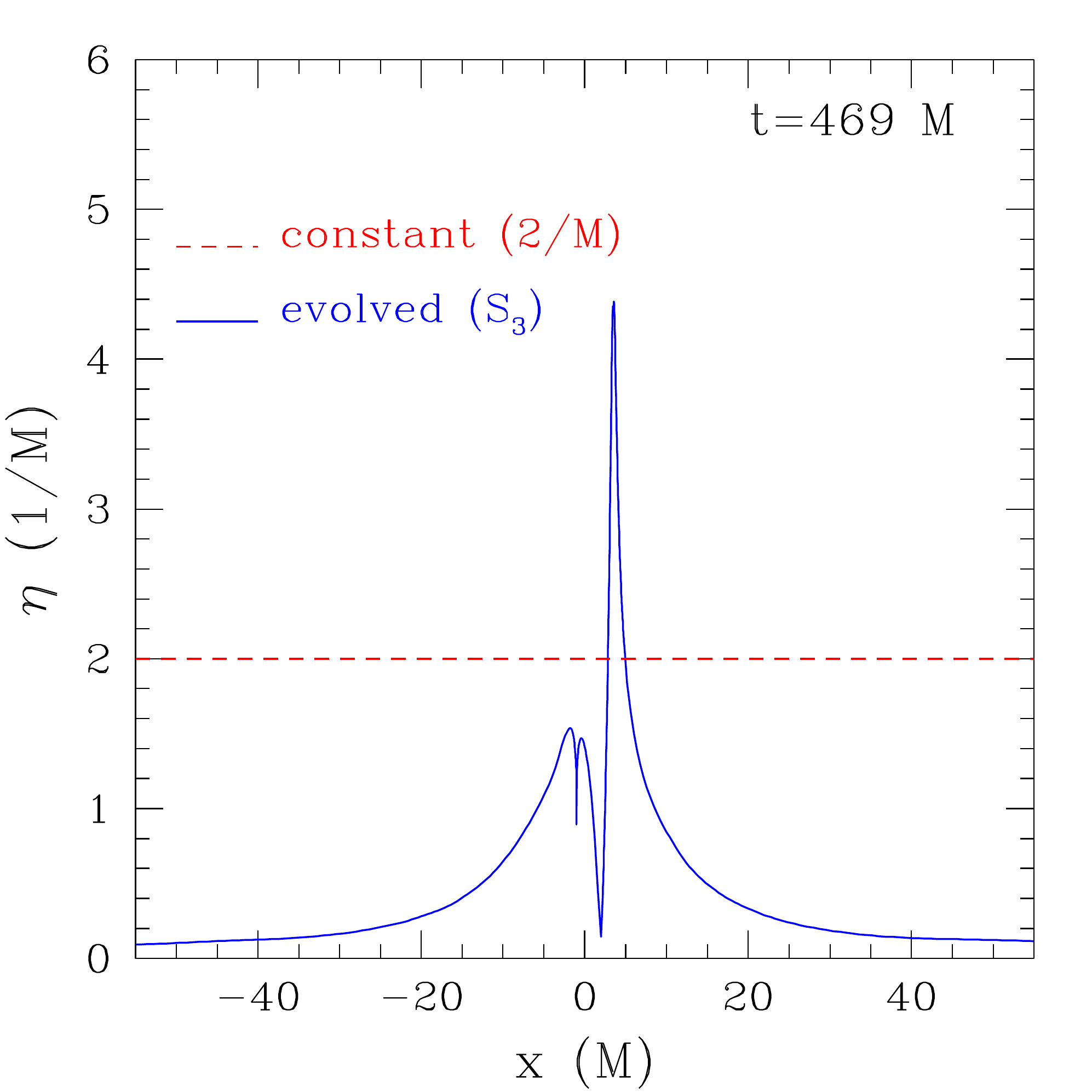}
\vskip 1.0cm
\includegraphics[width=\textwidth]{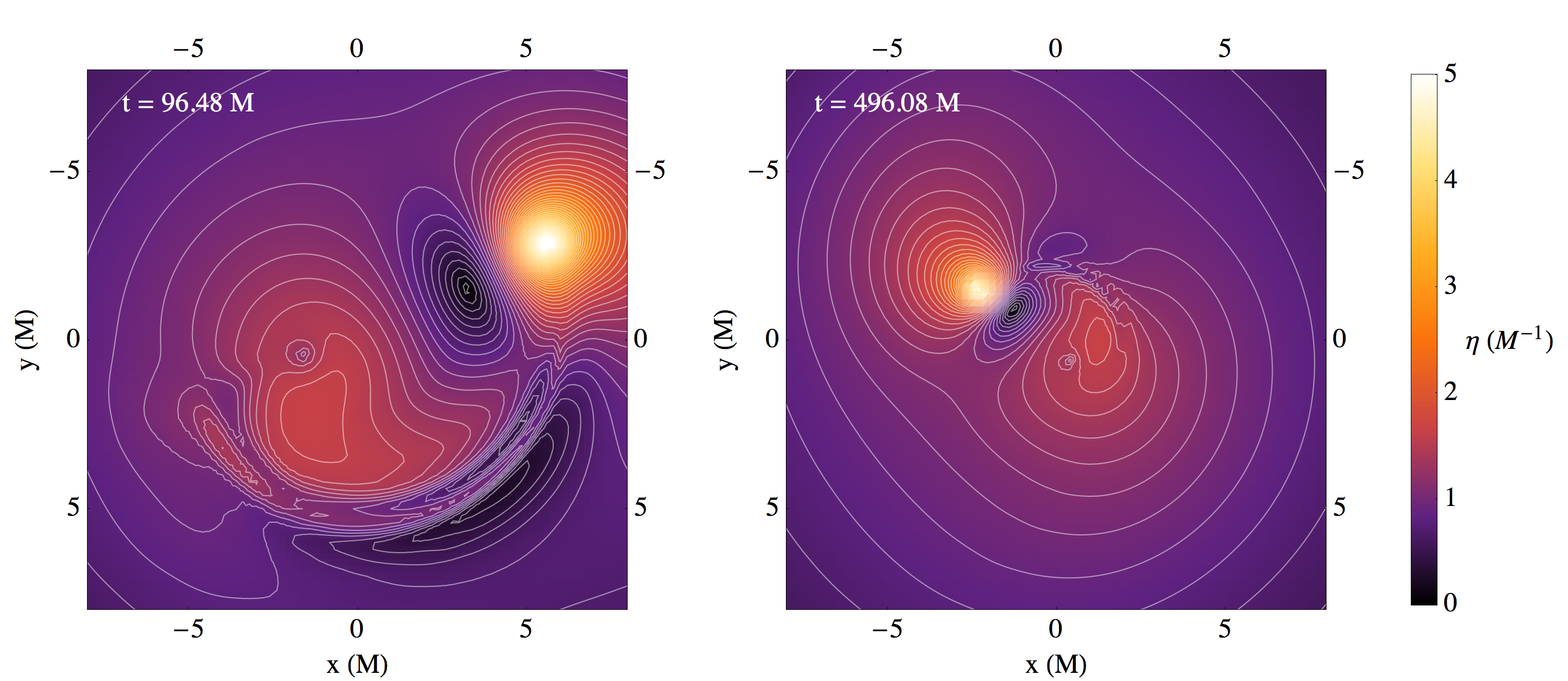}
\caption{Spatial dependence of the damping term $\eta$ for an
  \textit{unequal-mass} BBH system with $q=1/4$. Different lines refer
  to the case when $\eta$ is kept constant in space and time (red
  dashed line) or when it is evolved in time using
  Eq. (\ref{DampEvol}) with source given by (\ref{DampE3}) (blue solid
  line). The top left and right panels refer to two different times
  when the punctures are close to the $x$-axis and show that the new
  gauge drives the damping term to follow the BHs and be smooth
  elsewhere. The bottom left and right panels, instead, show the
  damping term on the $(x,y)$ plane at two representative times and
  show the non trivial but smooth distribution of the damping term
  which adapts to the motion of the punctures (the latter are near the
  maxima of the $\eta$).
}
\label{q4D8}
\end{figure*}

It is clear that when using either expression for the source
function, the value of $\eta$ at the location of the punctures adapts
in time through the coupling with the conformal factor $W$, which
tracks the position and the masses of the two BHs. It is also worth
noting that near the two punctures, the two evolution equations yield
very similar solutions for the damping factor as one would expect
since $R_0 \gg D$. However, when using the source (\ref{SourceD1})
(red dashed line), the evolution of the damping term produces large
gauge pulses which travel outwards and are amplified by the
mesh-refinement and the interpatch boundaries. This is particularly
evident in the right panel of Fig.~\ref{evolBBH} which refers to a
later time when the BHs have already merged. These undesirable
features are very effectively removed by adopting the source term
(\ref{SourceED1}) (blue solid line), which provides a natural fall-off
for the damping term as this propagates towards the outer
boundary.

\subsection{Unequal-mass binaries}

We next consider the evolution of an unequal-mass nonspinning BH
binary with mass ratio $q=1/4$, initial separation $D=8\,M$ and which
performs about $4$ orbits before merging and settling to an isolated
spinning BH after about $600\,M$. A complete list of the binary's
properties can be found in Table~\ref{tab:configs}.  We find that an
evolution equation for $\eta$ is a very convenient choice also for
unequal mass binary simulations.

\subsubsection{Spatial Dependence}

Fig.~\ref{q4D8} shows the spatial dependence of the damping
term at some representative times. The two top panels
refer to when the punctures are close to the $x$-axis for the cases where
$\eta$ is kept constant in space and time (red dashed line) and where it
is evolved in time with a source given by (\ref{DampE3}) (blue solid
line). Clearly, also in the unequal-mass case the new gauge drives the
damping term to follow the BHs and to be smooth elsewhere. The bottom
left and right panels show the damping term on the $(x,y)$
plane at two times which are similar to those shown in the top panels
and show the non-trivial but smooth distribution of the damping term which adapts to
the motion of the punctures (the latter are near the maxima of the
$\eta$).
The value of $\eta$ is $\sim 1.5/M$ near the large BH and $\sim 4.5/M$
near the small BH, leading to a value for $m_i \, \eta \sim 1$ for
both BHs, where $m_1 = 0.8\, M$ and $m_2 = 0.2\, M$ are the
irreducible masses of each BH.  This demonstrates that the gauge
condition is adapting the value of $\eta$ to the mass of the BH.  As
an aside, we note that there is a region between the BHs where $\eta$
drops nearly to 0, and a ``wake'' of low $\eta$ which follows behind
the motion of the smaller BH.  

\subsubsection{Waveforms and Errors}

\begin{figure}
\includegraphics[width=8.0cm]{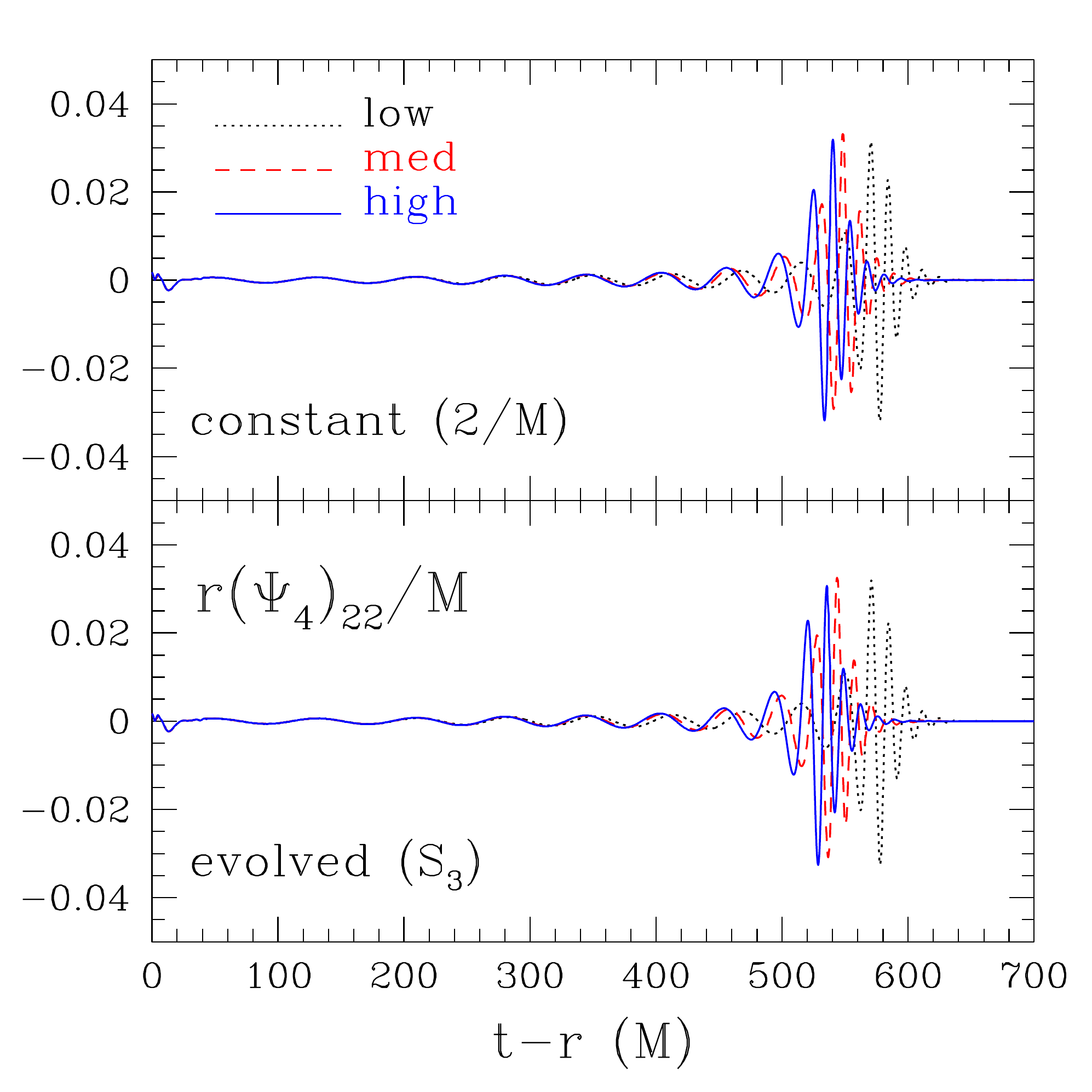}
\caption{Real part of the $\ell = 2, m = 2$ mode of the gravitational
  waveform $\Psi_4$ for the unequal-mass black-hole binary. Different
  lines refer to different resolutions (see text for details) and the
  two panels refer to the case when $\eta$ is kept constant in space
  and time (upper panel) or when it is evolved with source given by
  (\ref{DampE3}) (lower panel).}
\label{fig:psi4re}
\end{figure}

\begin{figure*}
\includegraphics[width=8.0cm]{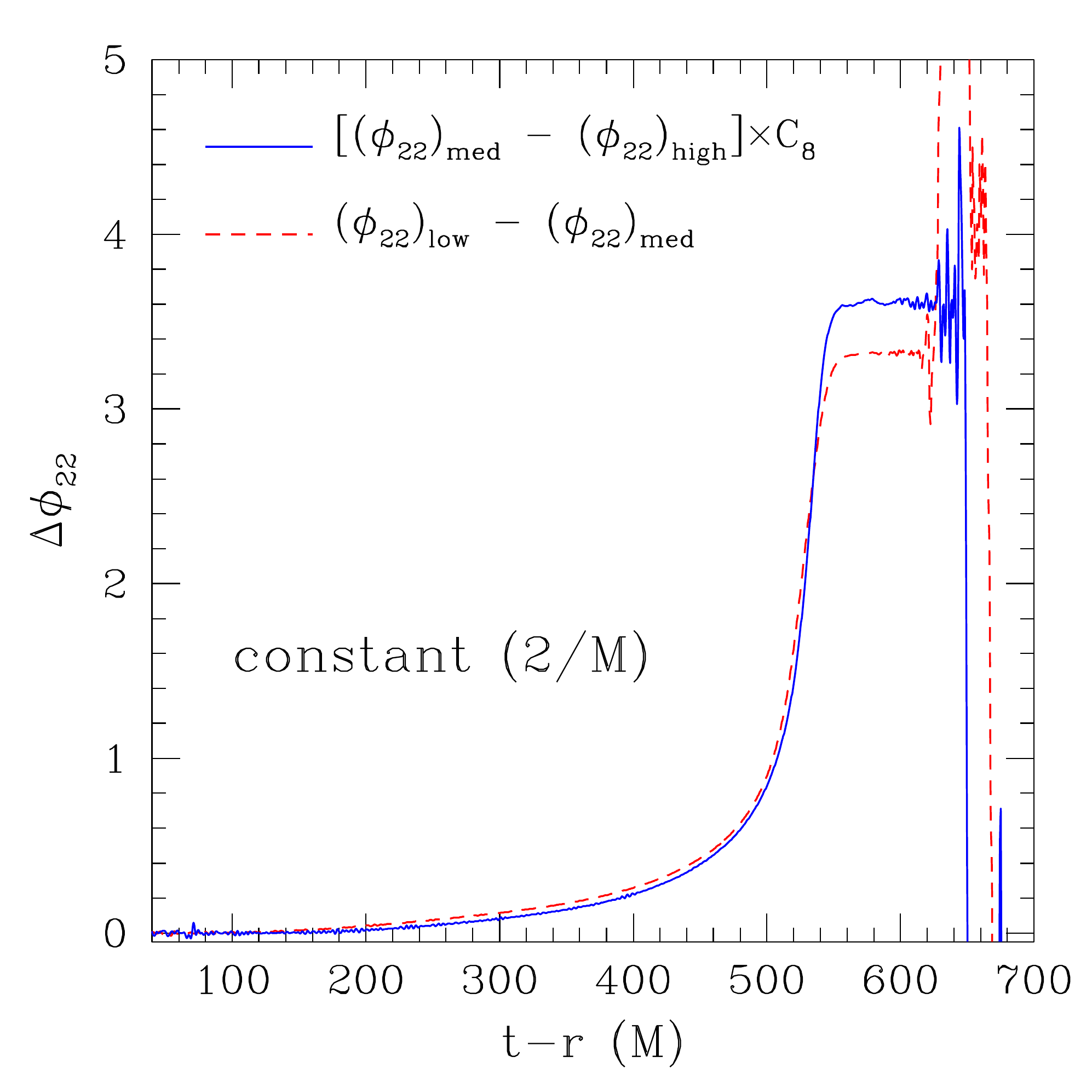}
\hskip 1.0cm
\includegraphics[width=8.0cm]{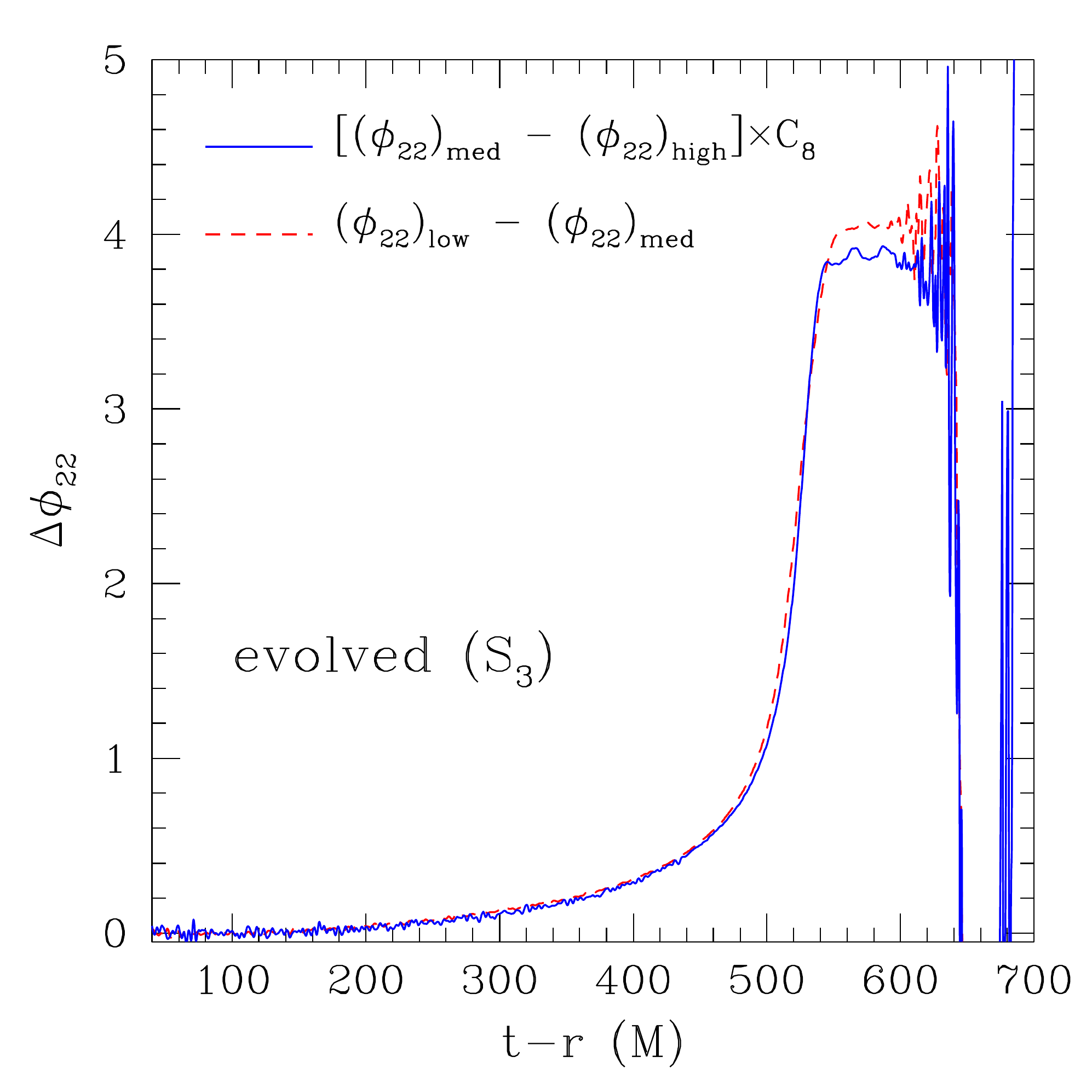}
\caption{Phase difference in the $\ell = 2, m = 2$ mode of the
  gravitational waveforms as computed between the low and medium
  resolutions (red dashed line) and between the medium and high
  resolutions (blue solid line). The latter has been scaled to
  compensate for an $8$th-order convergence which is indeed obtained
  as shown by the good overlap between the curves. The left panel
  refers to the case when $\eta$ is kept constant in space and time,
  while the right one to when it is evolved with source given by
  (\ref{DampE3}).}
\label{fig:psi4reserr}
\end{figure*}

One of the most useful quantities calculated from a BBH simulation is
the gravitational waveform.  The physical waveform is that measured at
future null infinity, and is independent of the gauge condition used
for the evolution.  In practice, waveforms are often computed at
finite radii and extrapolated to future null infinity, and the gauge
condition can have an effect on the extrapolation error.  More
importantly, different gauges can lead to different truncation errors
in the evolution of the BH motion depending on how well we can
reproduce a given field variable at the chosen resolution, and these
errors will be reflected in the waveform errors.  As a result,
different gauges can in practice lead to small differences also in the
calculation of the waveforms, and it is important to study the impact
of the different gauges on the waveforms and their truncation errors.

In Fig.~\ref{fig:psi4re} we show the real part of the $\ell = 2,
m = 2$ mode of the gravitational waveform $\Psi_4$ as extracted at $r
= 100\,M$. Different lines (dotted, dashed, solid) refer to the
different resolutions reported in Table~\ref{tab:grid}, while the two
panels refer to the cases where $\eta$ is kept constant in space and
time (upper panel) and where it is evolved with a source given by $S_3$
[\cf Eq.~(\ref{DampE3})] (lower panel). As expected, the differences in
the waveforms due to
the different truncation errors for the two gauges are very small and
confirm that a different prescription for the gauges does not
influence the gravitational-radiation
signal. Figure~\ref{fig:psi4reserr} offers a different view of
the gravitational-wave signal by showing the phase difference in the
$\ell = 2, m = 2$ mode, $\Delta \phi_{22}$, of the waveforms as
computed between the low and medium resolutions (red dashed line) and
between the medium and high resolutions (blue solid line). The latter
has been scaled to compensate for an $8$th-order convergence which is
indeed obtained as shown by the very good overlap between the dashed
and solid curves during all of the inspiral, being slightly worse
during the merger, when the convergence order drops. The left panel
refers to the case when $\eta$ is kept constant in space and time,
while the right one refers to when it is evolved with a source given by
(\ref{DampE3}) and $R=20\,M$. Once again the phase errors are
comparable between the two gauge conditions.

We see that adopting an
evolution equation for the damping term allows one to reproduce with a
comparable accuracy the numerical results obtained with the more
standard prescription of a constant value for $\eta$. At the same
time, however, it also shows that in this way no special tuning is
required and the prescription that works well for equal-mass binaries
is also very effective for an unequal-mass case with $q=1/4$. We
expect this to be true also for much smaller mass ratios, whose
investigation goes beyond the scope of this paper but will be pursued
in our future work.

\subsubsection{Impact on the Apparent-Horizon Size}

\begin{figure*}
\includegraphics[width=8.0cm]{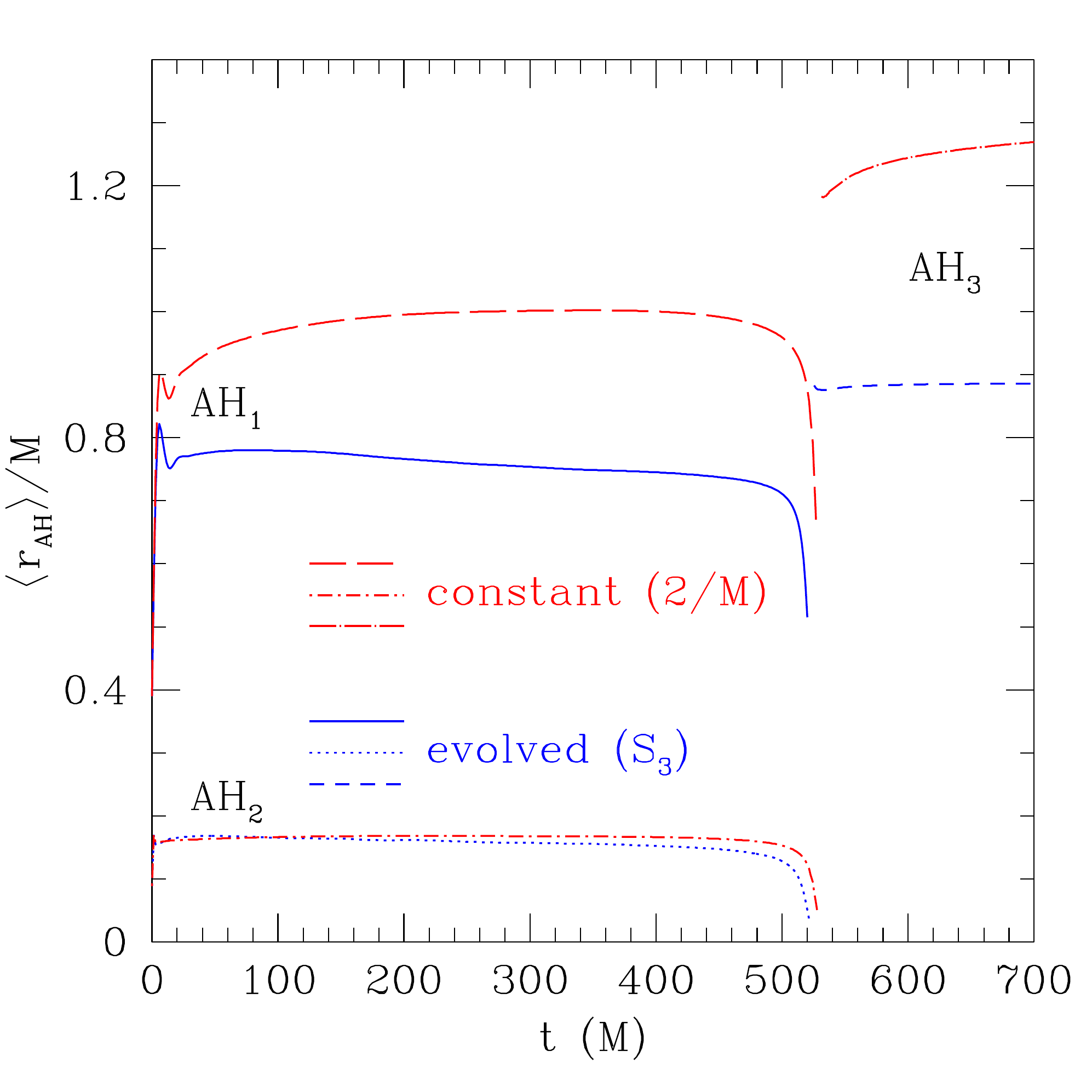}
\hskip 1.0cm
\includegraphics[width=8.0cm]{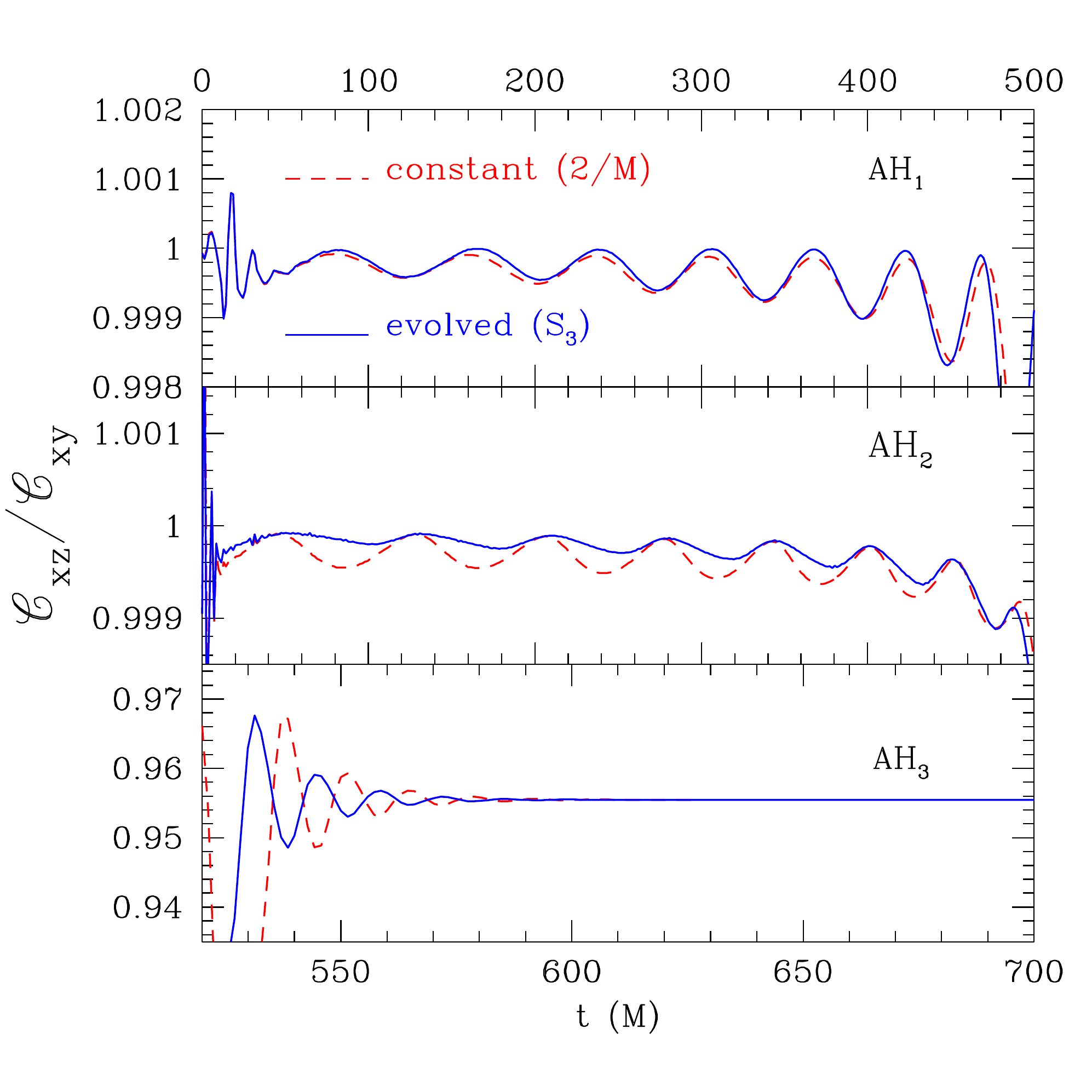}
\caption{\textit{Left panel:} Evolution of the average radius of the
  AHs on the $(x,y)$ plane before and after the merger. Shown with red
  long-dashed (large BH), dot-dashed (small BH) and dashed lines
  (merged BH) are the radii obtained for the case in which $\eta$ is
  kept constant in space and time. Shown instead with blue solid
  (large BH), dotted (small BH) and short-dashed lines (merged BH) are
  the radii obtained for the case in which $\eta$ is evolved with
  source given by (\ref{DampE3}). \textit{Right panel:} Evolution of
  the ratio of the AHs' proper circumferences on the $(x,z)$ and
  $(x,y)$ planes, ${\cal C}_{xy}$, ${\cal C}_{xz}$ as computed for the
  three BHs. The first two panels from the top refer to the AHs of the
  binary, while the bottom one to refers to the merged AH and thus
  covers a different range in time.}
\label{fig:fig7}
\end{figure*}

Some final consideration will now be given to the effect that the new
gauge condition has on the apparent horizon (AH) coordinate size as it varies during the
simulation. We recall that numerical simulations of moving punctures
have routinely reported a certain dynamics in the coordinate size
of the AH. This dynamics takes place during the early stages of the
evolution, as the gauges evolve rapidly from their initial conditions and
reach the values they are brought to by the time-dependent
drivers. Although these changes are of purely gauge nature and do not
influence the subsequent evolution of gauge-invariant quantities, they
represent nevertheless a computational nuisance as they require a
high-resolution mesh-refinement box (\ie the one containing the BH) to
be sufficiently large so as to accommodate the AH as it grows. Clearly,
this requirement becomes particularly important for binaries with very
small ratios, as in this case it is desirable to have the least
dynamics and thus reduce the computational costs.

To report the ability of the new gauge to reduce the variations in
the AH size, we show in the left panel of Fig.~\ref{fig:fig7} the time
evolution of the average radii of the unequal-mass binary on the
$(x,y)$ plane before and after the merger. We recall that during the
inspiral and merger we follow three different AHs, two corresponding
to the initial BHs (\ie ${\rm AH}_1$ and ${\rm AH}_2$) and a third
one, which is produced at the merger and contains the first two
(\ie ${\rm AH}_3$). We also note that the two initial AHs can still be
followed for a certain amount of time after a single AH is found
comprising the two. Shown in the left panel of Fig.~\ref{fig:fig7}
with red long-dashed (large BH), dot-dashed (small BH) and dashed
lines (merged BH) are the radii obtained for the case in which $\eta$
is kept constant in space and time.  The radii obtained when $\eta$ is
evolved with source given by (\ref{DampE3}) are shown with blue solid
(large BH), dotted (small BH) and short-dashed lines (merged BH).

In each case, the individual AHs first grow in radius as they
initially adapt to the chosen gauge, \ie for $t\lesssim 20\,M$. After
this initial stage, however, the two gauges show a different behaviour
and in the new gauge the AHs tend to maintain their coordinate size
more closely than in the constant-$\eta$ case (\cf blue solid and red
long-dashed lines).  This ability to conserve the original size is
evident also after the formation of the common AH at $t \simeq
520\,M$, where the new gauge shows an average radius for AH$_3$ which
is essentially constant for $t \sim 200\,M$, while it grows of about
$10\%$ over the same timescale when using a constant $\eta$
gauge. These effects provide benefits for the
mesh-refinement-treatment in our BBH simulations. Namely, they allow
us to reduce the extent of the finest mesh-refinement levels
containing the AHs and reduce therefore the computational cost of the
simulations while keeping a given resolution.

Besides the changes in the overall coordinate size of the AHs
discussed above, it is interesting to consider how much their shape
changes in time in the two gauges and this is summarised in the right
panel of Fig.~\ref{fig:fig7}, where we show the evolution of the ratio
of the AHs' proper circumferences on the $(x,z)$ and $(x,y)$ planes,
${\cal C}_{xz}/{\cal C}_{xz}$, as computed for the three BHs (the
first two panels from the top refer to the AHs of the binary, while
the third one to the merged AH and thus a different time
range). Overall, in both gauges the merging BHs remain spherical to a
few parts per thousand.  We have also verified that the masses of the BHs
as computed from the AHs are consistent between the two gauges within numerical errors. Interestingly, the new gauge leads
to smaller oscillations in the ratio for the smallest of the BHs. This
is a very small improvement, which however minimises spurious
gauge-dynamics and helps when making AH-based measurements.

\section{Conclusions}
\label{conclusions}

Even with a complete computational infrastructure,
numerical-relativity simulations of inspiralling compact binaries
would not be possible without suitable gauge conditions. A large bulk
of work developed over the last decade has provided gauge conditions
for the lapse and the shift which have been used with success both in
vacuum and non-vacuum spacetimes when simulating binaries with
comparable masses. However, as the need to investigate black-hole
binaries with small mass ratios increases, evidence has emerged that
the standard ``Gamma-driver'' shift condition requires a careful and
non-trivial tuning of its parameters to ensure long-term stable
evolutions of such binaries.

As a result, a few different suggestions have been made recently in
the literature to improve the Gamma-driver condition for the shift and
these have focused, in particular, on the specification of a
spatially dependent damping term $\eta$. This approach has been shown
to work well under some conditions but not always and prescriptions
which are effective in some cases can lead to instabilities in others.
In addition, the prescriptions require the specification of
coefficients whose tuning may be dependent on the mass ratio in a way
which is not trivial.

Following a different approach, we have presented a novel gauge
condition in which the damping constant is promoted to be a dynamical
variable and the solution of an evolution equation. We show that this
choice removes the need for special tuning and provides a shift
damping term which is free of instabilities for all of the spacetimes
considered. Although rather trivial, our gauge condition has a number
of advantages: \textit{i)} it is very simple to implement numerically
as it has the same structure of the other gauge conditions;
\textit{ii)} it adapts dynamically to the individual positions and
masses of the BBH system and could therefore be used also for binaries
with very small mass ratios; \textit{iii)} it reduces the variations
in the coordinate size of the apparent horizon of the larger black
hole thus limiting the computational costs; \textit{iv)} all of the
complexity in the new gauge is contained in the source function which
can be easily improved further.  This last point will be part of our
future research in this direction.

\acknowledgments

It is a pleasure to thank Jose-Luis Jaramillo for useful
discussions. This work was supported in part by the DFG grant
SFB/Transregio~7 ``Gravitational-Wave Astronomy''.  The computations
were performed on the Damiana cluster at the AEI, at LRZ Munich, on
the LONI network (\texttt{www.loni.org}), and on the Teragrid network
(allocation TG-MCA02N014).

\appendix

\bibliographystyle{apsrev-nourl-noeprint}
\bibliography{aeireferences}

\end{document}